\documentclass[preprint,showpacs,preprintnumbers,nofootinbib,amsmath,amssymb,aps,prc,longbibliography]{revtex4-1}
\usepackage{graphicx}
\usepackage{dcolumn}
\usepackage{bm}
\usepackage{amsmath}
\usepackage{mathtools}
\begin{document}
\title{Effective restoration of dipole sum rules within the renormalized random-phase approximation}
\author{N. Quang Hung$^{1}$}
 \email{nqhungdtu@gmail.com}
 \author{N. Dinh Dang$^{2}$}
  \email{dang@riken.jp}
 \author{T.V. Nhan Hao$^{1,3}$}
 \email{On leave of absence from the Center for Theoretical and Computational Physics, College of
Education, Hue University, Hue City, Vietnam; tran.viet.nhan.hao@gmail.com}
 \author{L. Tan Phuc$^{1}$}
 \email{letanphuc191190@gmail.com}
 \affiliation{1) Institute of Research and Development, Duy Tan University,  K7/25 Quang Trung, Danang city, Vietnam\\
 2) Quantum Hadron Physics Laboratory, RIKEN Nishina Center for Accelerator-Based Science, 2-1 Hirosawa, Wako City, 351-0198 Saitama, Japan\\
and Institute for Nuclear Science and Technique, Hanoi, Vietnam\\
 3) Department of Physics and Astronomy, Texas A$\&$M University-Commerce, Commerce, TX 75429, USA\\
}

\date{\today}
\begin{abstract}
The dipole excitations for calcium and zirconium isotopes are studied within the fully self-consistent Hartree-Fock mean field incorporated with the renormalized random-phase approximation (RRPA) using the Skyrme interaction SLy5. The RRPA takes into account the effect of ground-state correlations beyond RPA owing to the Pauli principle between the particle-hole pairs that form the RPA excitations as well as the correlations due to the particle-particle and hole-hole transitions, whose effects are treated here in an effective way. By comparing the RPA results with the RRPA ones, which are obtained for isoscalar (IS) and isovector (IV) dipole excitations in $^{48, 52, 58}$Ca and $^{90, 96, 110}$Zr, it is shown that ground-state correlations beyond the RPA reduce the IS transition strengths. They also shift up the energy of the lowest IV dipole state and slightly push down the peak energy of the IV giant dipole resonance. As the result, the energy-weighted sums of strengths of both IS and IV modes decrease, causing the violation of the corresponding energy-weight sum rules (EWSR). It is shown that this sum rule violation can be eliminated by taking into account the contribution of the particle-particle and hole-hole excitations together with the particle-hole ones in a simple and perturbative way. Consequently, the ratio of the energy-weighted sum of strengths of the pygmy dipole resonance to that of the giant dipole resonance increases.
\end{abstract}

\maketitle
\section{Introduction}
\label{Intro}
The random-phase approximation (RPA) is one of the most frequently used methods in the theoretical study of vibrational collective excitations in low-energy physics. These collective modes include the low-lying vibrations and high-lying giant resonances. Within the RPA, these vibrational excitations are generated by the coherent superpositions of elementary excitations. For instance, the electric multipole excitations $EJ^{\pi}$ ($J^{\pi} = 0^{+}, 1^{-}, 2^{+}, 3^{-}$, etc.), including the isoscalar (IS) and isovector (IV) giant resonances, consist of many particle-hole (ph) components. 

The basic assumption of the RPA is the quasi-boson approximation (QBA), which implies that the operators of ph pairs behave like ideal bosons, ignoring the Pauli principle owing to their fermion structure. Algebraically, the QBA is the replacement of the exact commutation relation between the ph pair operators $[B_{ph}, B^{\dagger}_{p'h'}]$ with its expectation value $\langle HF|[B_{ph}, B^{\dagger}_{p'h'}]|HF\rangle$ in the Hartree-Fock (HF) ground state, where $B_{ph}^{\dagger}= a^{\dagger}_pa_h$ with $a^{\dagger}_p$ and $a_h$ being the particle (p) creation and hole (h) annihilation operators, respectively~\cite{Ring}. This assumption perfectly holds when the RPA vibrations consist of many coherent ph excitations, for instance in the case of high-lying collective excitations such as the giant resonances or low-lying ones such as the first $2^{+}$ or $3^{-}$ states in medium and heavy nuclei, where the concept of a nuclear mean field works very well.  However, in light nuclei the validity of the mean-field concept and the QBA itself become questionable. Hence it becomes necessary to take into account properly the fermion structure of the ph pairs, removing in this way the violation of the Pauli principle. 

There have been several methods to preserve the Pauli principle between the ph pairs by including the correlations in the ground state, which are neglected within the QBA. These correlations are referred to as the ground-state correlations (GSC) beyond RPA.  Among the earliest approaches to include GSC beyond RPA, the one proposed by Hara~\cite{Hara} is rather transparent and simple because it evaluates the expectation value $\langle 0|[B_{ph}, B^{\dagger}_{p'h'}]|0\rangle$ of the commutation relation $[B_{ph}, B^{\dagger}_{p'h'}]$ in the correlated RPA ground state $|0\rangle$ by using the diagonal approximation, resulting in the particle and hole occupation numbers, $f_k\equiv \langle 0|a^{\dagger}_ka_k|0\rangle$ ($k = p, h$). The latter are approximately expressed in terms of the RPA backward-going amplitudes $Y_{ph}^{\nu}$. This leads to the renormalized RPA (RRPA) equation, nonlinear with respect to the amplitudes $Y_{ph}^{\nu}$, which can be solved by iteration. In the RRPA equation the residual interactions are renormalized by the so-called GSC factor $D_{ph}\equiv \sqrt{f_h- f_p}$, which is a function of $(Y_{ph}^{\nu})^2$. 

In a later development, Catara, Dang and Sambataro  derived in Ref. \cite{CaDaSa} an equation for the GSC factor $D_{ph}$, which recovers the expression for $D_{ph}^{Hara}$, which was obtained by Hara in the limit of small $Y_{ph}^{\nu}$. It has been shown by Dukelsky and Schuck that this equation produces the exact expression for the GSC factor in the equidistant multilevel pairing model~\cite{Dukelsky}. However, the application of the RRPA using this GSC factor $D_{ph}$ has been limited so far only to the energy and $B(E3)$ value of the lowest $3^{-}_1$ state in $^{146}$Gd and $^{208}$Pb~\cite{CaDaSa}.  The RRPA was also studied by using the equations of motion method in Ref. \cite{Rowe} and the Green's functions techniques in Ref. \cite{Schuck}. An alternative derivation of the occupation numbers by using the number-operator method was carried out in Refs. \cite{Rowe,Lenske}. This method was employed in Ref. \cite{Catara} to obtain an expression for the GSC factor $D_{ph}$ within the improved RPA up to the order of $O(Y^{4})$, which was applied to the calculations in metallic clusters.

The results of Ref. \cite{CaDaSa} show that, due to the GSC beyond the RPA, the energy the $3^{-}_1$ state in these two nuclei increases upward, leading to the significant decrease of around 40$\%$ and 20$\%$ in the associate $B(E3)$ values in $^{146}$Gd and $^{208}$Pb, respectively. These effects are more pronounced than those predicted for the low-lying $2^+$ states~\cite{Hara}. No investigation was carried out for the $1^-$ states, whose first moment $m_1$, known as the energy-weighted sum rule (EWSR), is the most important. For the $E1$ transitions, this EWSR is related to the model-independent Thomas-Reich-Kuhn sum rule TRK = 60 NZ/A (MeV mb) for the giant dipole resonance (GDR) in a nucleus with A nucleons (N neutrons and Z protons).  When a velocity-dependent effective interaction, such as the Skyrme one, or the meson-exchange interaction is used, the velocity-dependent and/or exchange terms contribute to the IV multipole sum rules. The IV $E1$ EWSR then exceeds the TRK as TRK$\times(1+ {\kappa})$ with 0$ <{\kappa}<$ 1~\cite{Liu}. The RPA (with the velocity-independent interaction) fulfills the GDR sum rule. However, being multiplied by the factor $D_{ph}$, the RRPA matrix elements of the ph interactions are reduced. This leads not only to a shift down of the GDR energy but also to a decrease in the $B(E1)$ values, causing the violation of the GDR sum rule.

The problem of model-independent sum rule violation within the RRPA was first pointed out in Ref. \cite{Hirsch}, where the authors attributed the origin of the problem to the adopted structure of both the Hamiltonian and the transition operators, which neglect the scattering terms coming from the particle-particle (pp) and hole-hole (hh) transitions. In Ref. \cite{Catara}, it has been shown that the improved RPA does not fulfill the EWSR as well. A practical idea to restore the EWSR within the RRPA was suggested in the same paper in the same spirit of Ref. \cite{Hirsch} by noticing that the transition operator, in general has not only a ph part but also the particle-particle (pp) and hole-hole (hh) ones. Therefore, in order to restore the EWSR, the RRPA transition amplitudes should contain also the contributions from the pp and hh transitions.

As a matter of fact, the importance of the pp and hh transitions were evident already in the studies of the finite-temperature RPA (FT-RPA) in the 1970s - 1980s, where the FT-RPA equations were derived and studied in details ~\cite{Ignatyuk,Sommer,Vautherin,DangJP}. The FT-RPA equations, which were derived in these papers for the separable interactions, are formally identical to the RRPA ones if the ph transitions are considered. The only difference is that the single occupation numbers $f_k$ within the FT-RPA are described by the Fermi-Dirac distributions for non-interacting fermions at finite temperature $T$. It has been shown in Ref. \cite{Sommer}, that the GDR energy decreases with increasing $T$ for the ph transitions within the FT-RPA, because $D_{ph}$ becomes smaller than 1 and decreases with increasing $T$. However, after the pp and hh transitions are included together with the ph ones, the GDR energy remains nearly the same as its value when $T=$ 0 at $T$ varies up to 4 MeV. The $E2$ and $E3$ ESWRs also remain essentially constant up to $T=$ 2 MeV. As for the IV $E1$ excitations with thin the FT-RPA, the results of calculations obtained for $^{40}$Ca, including ph as well as pp and hh transitions by using a zero-range force in Ref. \cite{Vautherin} have also shown that the EWSR remains nearly constant up to temperature $T=$  6 MeV. A similar conclusion was obtained in Ref. \cite{Lacroix}, where the Skyrme force SGII was used in calculations.

With increasing the number of neutrons, a phenomenon called pygmy dipole resonance (PDR) is predicted in neutron-rich nuclei and a number of theoretical and experimental studies were devoted to its properties in the recent years \cite{PDRexp,ReRPA,Tarpanov,Sarchi,SRPA,ReRPA1,QPM1,QPM2}. The PDR has often been interpreted as a manifestation of the oscillation of the weakly-bound neutron skin against the isospin-symmetric core of protons and neutrons, although this picture is rather sketchy. As a matter of fact, the nature of the PDR still remains an issue open to debate. Different theoretical models still do not agree on its collectivity and coherence. For example, while the relativistic random-phase approximation predicted a prominent peak below 10 MeV in $^{120,132}$Sn and $^{122}$Zr~\cite{ReRPA}, which is identified as the collective PDR, the results of calculations including monopole pairing within the quasiparticle RPA (QRPA) show only a group of slightly collective states \cite{Tarpanov} in $^{122,130}$Sn, whereas for $^{120,132}$Sn~\cite{Sarchi} no collective $E1$ states are even seen in the low-energy region. As the PDR is located in a low energy region, the effect of GSC beyond the RPA, which are included in the phRRPA, should be taken into account in the study of the PDR. This study should be complementary to those, which include 2p-2h excitations such as the second RPA~\cite{SRPA}, the relativistic RPA plus phonon-coupling model~\cite{ReRPA1}, the quasiparticle-phonon model~\cite{QPM1}, a version of which includes even couplings upto 3p3h configurations in terms of three-phonon components~\cite{QPM2}. However, the major drawback, as mentioned above as well as in Ref. \cite{Gamba}, is that all extensions of the RPA to the RRPA so far violate the energy-weighted sum rules (EWSRs) for IS and IV dipole excitations. 

Obviously, in order to restore the $E1$ sum rule, one should derive and solve the complete set of RRPA equations that include all the ph, as well as pp and hh configurations. The transition probabilities, which are calculated based on these solutions, should also include all ph, pp, and hh transitions. This is  undoubtedly a formidable task, which we are now working on. In the present paper, as a preparatory study, we would like to see if the problem can be resolved in an approximate and simple way, which can nonetheless be applied in practical calculations of electric dipole excitations. To this end, we modify the computer code for self-consistent RPA with Skyrme-type interactions, which has been developed by Col\`{o} and collaborators~\cite{Colo}, to include the effect of GSC beyond the RPA following the method proposed in Ref. \cite{CaDaSa}. The dipole EWSRs are restored by perturbatively taking into account also the contribution of pp and hhole configurations, which have been neglected so far in all practical calculations within the RRPA. 

The paper is organized as follows. The formalism of the RRPA is summarized in Secs. \ref{secRRPA} and \ref{fRRPA}. The analysis of numerical calculations for the pigmy dipole and giant dipole excitations in $^{48-58}$Ca and $^{90-110}$Zr are presented in Sec. \ref{results}. The paper is summarized in the last section, where conclusions are drawn.

\section{The RRPA for particle-hole vibrations ($\mathrm{ph}$RRPA)}
\label{secRRPA}
The renormalized RPA for particle-hole vibrations, which is referred to as the phRRPA, has been discussed in details in Refs. \cite{Hara,CaDaSa,Dukelsky,Rowe,Schuck,Lenske, Catara}. Therefore, we summarize here only its main formulation within the ph representation including all angular momentum couplings in spherical basis, which is necessary for its application to the calculations in the present paper.

The RPA operator $Q_{JMi}^{\dagger}$ is constructed as a superposition of ph-pair operators in the form~\cite{Ring}  
\begin{equation}
Q_{JMi}^{\dagger} = \sum_{ph}\bigg[X_{ph}^{Ji}B_{ph}^{\dagger}(JM)- Y_{ph}^{Ji}B_{ph}(J\overline{M})\bigg]~,
\label{Q}
\end{equation}
where $B_{ph}^{\dagger}(JM)$ is the ph-pair creation operator
\begin{equation}
B_{ph}^{\dagger}(JM) = \sum_{m_pm_h}\langle j_pm_p j_hm_h|JM\rangle a^{\dagger}_{j_pm_p}a_{j_h\overline{m}_h}~,
\label{Bph}
\end{equation}
which couples the particle creation operator $a_{j_pm_p}^{\dagger}$ on the orbital having the angular momentum $j_p$ and projection $m_p$ with the
hole annihilation operator $a_{j_h {-m}_h}$ having the angular momentum $j_h$ and projection $-m_h$, where $m_k$ (with $k= p, h$) takes $(2j_k+1)$ values, namely $m_k = -j_k, -j_k +1, ..., j_k -1, j_k$. The notations 
$a_{j\overline{m}}\equiv (-)^{j+m}a_{j-m}$ and $B_{ph}(J\overline{M})\equiv(-)^{J+M}B_{ph}(J-M)$ are used for time-reversal operators. The quantum numbers $J$ and $M$ ($M = -J, -J+1,..., J-1, J$) denote the total angular momentum (the multipolarity of the phonon excitation) and its projection, respectively.

The RPA excited states are defined as
\begin{equation}
|JMi\rangle = Q_{JMi}^{\dagger}|0\rangle~,
\label{JMi}
\end{equation}
where the RPA correlated ground state $|0\rangle$ is the vacuum of the phonon operator, that is
\begin{equation}
Q_{JMi}|0\rangle = 0~,
\label{0}
\end{equation}
and the excited states should be mutually orthonormal, namely
\begin{equation}
\langle JMi|J'M'i'\rangle = \langle 0|[Q_{JMi},Q_{J'M'i'}^{\dagger}]|0\rangle = \delta_{JJ'}\delta_{MM'}\delta_{ii'}~.
\label{ortho}
\end{equation}

The ph-pair operators  $B_{ph}(JM)$ and $B_{ph}^{\dagger}(JM)$ satisfy the exact commutation relation
\[
[B_{ph}(JM),B_{p'h'}^{\dagger}(J'M')] = \delta_{j_pj_p'}\sum_{m_pm_hm_h'}\langle j_pm_pj_hm_h|JM\rangle
\langle j_pm_pj_h'm_h'|J'M'\rangle a^{\dagger}_{j_h\overline{m}_h}a_{j_h'\overline{m}_h'}
\]
\begin{equation}
- \delta_{j_hj_h'}\sum_{m_pm_p'm_h}\langle j_pm_pj_hm_h|JM\rangle
\langle j_p'm_p'j_hm_h|J'M'\rangle a^{\dagger}_{j_p'{m}_p'}a_{j_p{m}_p}~.
\label{BB}
\end{equation}

The expectation value of this commutation relation in the RPA correlated ground state, $|0\rangle$ 
can be approximated as
\[
\langle 0|[B_{ph}(JM),B_{p'h'}^{\dagger}(J'M')]|0\rangle =
\]
\begin{equation}
\delta_{j_pj_p'}\sum_{m_pm_hm_h'}\langle j_pm_pj_hm_h|JM\rangle
\langle j_pm_pj_h'm_h'|J'M'\rangle\langle 0| a^{\dagger}_{j_h\overline{m}_h}a_{j_h'\overline{m}_h'}|0\rangle
\label{<BB>}
\end{equation}
\[
- \delta_{j_hj_h'}\sum_{m_pm_p'm_h}\langle j_pm_pj_hm_h|JM\rangle
\langle j_p'm_p'j_hm_h|J'M'\rangle \langle 0|a^{\dagger}_{j_p'{m}_p'}a_{j_p{m}_p}|0\rangle\simeq\delta_{JJ'}\delta_{MM'}\delta_{j_pj_p'}\delta_{j_hj_h'}D_{ph}~,
\]
where the GSC factor $D_{ph}$ is defined as
\begin{equation}
 D_{ph} \equiv f_h - f_p = \langle 0|a^{\dagger}_{j_h{m}_h}a_{j_h{m}_h}|0\rangle - \langle 0|a^{\dagger}_{j_p{m}_p}a_{j_p{m}_p}|0\rangle~,
 \label{GSCDph}
 \end{equation}
with $f_k\equiv\langle 0|a^{\dagger}_{j_k{\pm m}_k}a_{j_k{\pm m}_k}|0\rangle$ being the particle and hole occupation numbers in the correlated ground state $|0\rangle$ for $k=p$ and $k=h$, respectively.
\subsection{The quasi-boson approximation (QBA)}
\label{secQBA}
Within the QBA, the expectation value (\ref{<BB>}) is replaced with that estimated in the HF ground state, $|HF\rangle$. This yields $D_{ph}^{HF}=$ 1 because $f_h^{HF}\equiv\langle HF|a^{\dagger}_{j_h{m}_h}a_{j_h{m}_h}|HF\rangle=$ 1 and $f_p^{HF}\equiv\langle HF|a^{\dagger}_{j_p{m}_p}a_{j_p{m}_p}|HF\rangle=$ 0, so that the expectation value of $[B_{ph}(JM),B_{p'h'}^{\dagger}(J'M')]$ in the HF ground state becomes
\begin{equation}
\langle HF|[B_{ph}(JM),B_{p'h'}^{\dagger}(J'M')]|HF\rangle =\delta_{JJ'}\delta_{MM'}\delta_{j_pj_p'}\delta_{j_hj_h'}~.
\label{QBA}
\end{equation}
This equation shows that, within the QBA, the ph-pair operators  $B_{ph}(JM)$ and $B_{ph}^{\dagger}(JM)$ behave like ideal boson operators, which mean that the following approximation holds
\begin{equation}
[B_{ph}(JM),B_{p'h'}^{\dagger}(J'M')]\simeq\delta_{JJ'}\delta_{MM'}\delta_{j_pj_p'}\delta_{j_hj_h'}~.
\label{QBA1}
\end{equation}
Within the QBA, the orthogonality condition (\ref{ortho}) becomes
\begin{equation}
\langle JMi|J'M'i'\rangle = \langle 0|[Q_{JMi},Q_{J'M'i'}^{\dagger}]|0\rangle\simeq\langle HF|[Q_{JMi},Q_{J'M'i'}^{\dagger}]|HF\rangle = \delta_{JJ'}\delta_{MM'}\delta_{ii'}~.
\label{QBAortho}
\end{equation}

The QBA (\ref{QBA}) and the orthogonormality condition (\ref{ortho}) between the RPA excited states lead to the following normalization condition for the amplitudes $X_{ph}^{Ji}$ and $Y_{ph}^{Ji}$ of the phonon operator (\ref{Q})
\begin{equation}
\sum_{ph}(X_{ph}^{Ji}X_{ph}^{J'i'} - Y_{ph}^{Ji}Y_{ph}^{J'i'}) = \delta_{JJ'}\delta_{ii'}~.
\label{norm}
\end{equation}
The closure relations 
\begin{equation}
\sum_{i}(X_{ph}^{Ji}X_{p'h'}^{Ji} - Y_{ph}^{Ji}Y_{p'h'}^{Ji}) = \delta_{pp'}\delta_{hh'}~,\hspace{5mm} 
\sum_{i}(X_{ph}^{Ji}Y_{p'h'}^{Ji} - Y_{ph}^{Ji}X_{p'h'}^{Ji}) = 0~,
\label{closure}
\end{equation}
ensure the following inverse expression of ph-pair creation operator $B_{ph}^{\dagger}(JM)$ in terms of phonon operators $Q_{J'M'i'}^{\dagger}$ and $Q_{JMi}$ 
\begin{equation}
B_{ph}^{\dagger}(JM) = \sum_{i}[X_{ph}^{Ji}Q_{JMi}^{\dagger} + Y_{ph}^{Ji}Q_{J\overline{M}i}]~.
\label{B}
\end{equation}

Within the RPA the occupation numbers $f_{k}$ are calculated by using the mappings~\cite{CaDaSa,Hage}
\begin{equation}
\sum_{m_p}a^{\dagger}_{j_pm_p}a_{j_pm_p}\rightarrow \sum_{JM}\sum_{h}B_{ph}^{\dagger}(JM)B_{ph}(JM)~,
\label{boson1}
\end{equation}
\begin{equation} 
\sum_{m_h}a_{j_hm_h} a^{\dagger}_{j_hm_h}\rightarrow\sum_{JM}\sum_{p}B_{ph}^{\dagger}(JM)B_{ph}(JM)~.
\label{boson2}
\end{equation}
These mappings are exact within the QBA, where the operators $B_{ph}^{\dagger}(JM)$ and $B_{ph}(JM)$ behave like ideal boson operators according to (\ref{QBA1}). In this case the commutation relations between  the right-hand sides of (\ref{boson1}) and (\ref{boson2}), and $B_{ph}^{\dagger}(JM)$ are  
\begin{equation} 
[\sum_{J'M'h'}B_{p'h'}^{\dagger}(J'M')B_{p'h'}(J'M'),B_{ph}^{\dagger}(JM)]=\delta_{j_pj_p'}B_{ph}^{\dagger}(JM)~,
\label{check1}
\end{equation}
\begin{equation} 
[\sum_{J'M'p'}B_{p'h'}^{\dagger}(J'M')B_{p'h'}(J'M'),B_{ph}^{\dagger}(JM)]=\delta_{j_hj_h'}B_{ph}^{\dagger}(JM)~,
\label{check2}
\end{equation}
which are exactly equal  to $[\sum_{m_p'}a^{\dagger}_{j_p'm_p'}a_{j_p'm_p'}, B_{ph}^{\dagger}(JM)]$ and 
$[\sum_{m_h'}a_{j_h'm_h'}a^{\dagger}_{j_h'm_h'}, B_{ph}^{\dagger}(JM)]$, respectively.

By using Eqs. (\ref{B}) to express the right-hand sides of Eqs. (\ref{boson1}) and (\ref{boson2}) in terms of phonon operators in combination with the property of phonon vacuum (\ref{0}) within the QBA, that is (\ref{QBAortho}), the occupation numbers $f_k$ are found within the QBA as~\cite{Rowe}
\[
f_p = \frac{1}{2j_p+1}\langle 0|\sum_{m_p}a^{\dagger}_{j_pm_p}a_{j_pm_p}|0\rangle=\frac{1}{2j_p+1}\sum_{Ji}(2J+1)\sum_h(Y_{ph}^{Ji})^2~,
\]
\begin{equation}
f_h = 1-\frac{1}{2j_h+1}\langle 0|\sum_{m_h}a_{j_hm_h}a^{\dagger}_{j_hm_h}|0\rangle = 1 - \frac{1}{2j_h+1}\sum_{Ji}(2J+1)\sum_p(Y_{ph}^{Ji})^2~.
\label{fpfh}
\end{equation}
\subsection{The phRRPA}
\label{phRRPA}
The phRRPA method proposes to use Eq. (\ref{<BB>}) rather than Eq. (\ref{QBA}), without imposing the QBA. Because the excited states should be orthonormal, this
leads to the renormalized phonon operators in the form~\cite{CaDaSa}
\begin{equation}
{\cal Q}_{JMi}^{\dagger} = \sum_{ph}\bigg[\frac{{\cal X}_{ph}^{Ji}}{\sqrt{D_{ph}}}{B}_{ph}^{\dagger}(JM)- \frac{{\cal Y}_{ph}^{Ji}}{\sqrt{D_{ph}}}{B}_{ph}(J\overline{M})\bigg]~,
\label{RQ}
\end{equation}
which strictly satisfy the orthonormality condition (\ref{ortho}) rather than (\ref{QBAortho}), in which ${\cal Q}^{\dagger}_{JMi}$ and 
${\cal Q}_{JMi}$ replace ${Q}^{\dagger}_{JMi}$ and 
${Q}_{JMi}$, retaining the same normalization condition (\ref{norm}) for the amplitudes ${\cal X}_{ph}^{Ji}$ and ${\cal Y}_{ph}^{Ji}$.

The inverse transformation of (\ref{RQ}) is
\begin{equation}
B_{ph}^{\dagger}(JM) = \sqrt{D_{ph}}\sum_{i}[{\cal X}_{ph}^{Ji}{\cal Q}_{JMi}^{\dagger} + {\cal Y}_{ph}^{Ji}{\cal Q}_{J\overline{M}i}]~,
\label{RB}
\end{equation}
instead of Eq. (\ref{B}).

By using Eqs. (\ref{<BB>}) and (\ref{RB}), as well as the mappings  (\ref{boson1}) and (\ref{boson2}), and proceeding similarly as has been done in Sec. \ref{secQBA}, the occupation numbers ${f}_k$ within the phRRPA are found as
\[
{f}_p = \frac{1}{2j_p+1}\langle 0|\sum_{m_p}a^{\dagger}_{j_pm_p}a_{j_pm_p}|0\rangle=\frac{1}{2j_p+1}\sum_{Ji}(2J+1)\sum_hD_{ph}({\cal Y}_{ph}^{Ji})^2~,
\]
\begin{equation}
{f}_h = \frac{1}{2j_h+1}\langle 0|\sum_{m_h}a^{\dagger}_{j_hm_h}a_{j_hm_h}|0\rangle = 1 - \frac{1}{2j_h+1}\sum_{Ji}(2J+1)\sum_pD_{ph}({\cal Y}_{ph}^{Ji})^2~.
\label{Rfpfh}
\end{equation}
It has been discussed in Ref. \cite{Rowe} that, if the number-operator method is used to derive the occupation numbers $f_k$, the expressions for $f_p$ and $f_h$ in Eq. \ref{Rfpfh} acquire a factor of 1/2 in front of $1/({2j_{p(h)}+1})$, which reduces the effect of GSC. However, as has been pointed out in Ref. \cite{Ellis}, the precise correspondence between the two methods is not immediately apparent, since it is not easy to pinpoint what effects are included in the number operator method. Moreover, the expressions obtained in Ref. \cite{Catara} for the same quantities include the term of even higher orders in ${\cal X}_{ph}^{Ji}$ and ${\cal Y}_{ph}^{Ji}$ amplitudes [See Eqs. (2.20) and (2.21) in Ref. \cite{Catara}], which makes them different from those given in Eq. (\ref{Rfpfh}) as well as in the number- operator method [Eq. (54) in Ref. \cite{Rowe}]. However, as has been already mentioned in Ref. \cite{Rowe}, in practice, such difference does not appear to be such a serious effect, since only a few low-lying collective excitations contribute coherently to the GSC. For many non-collective state the backward-going amplitudes ${\cal Y}_{ph}^{Ji}$ are small and and take place with random sign. Therefore, we decide to use the expressions in Eq. (\ref{Rfpfh}), keeping in mind that the resulting effect of GSC may be slightly overestimated compared to that predicted by the number-operator method. 

Notice that, within the phRRPA, Eqs. (\ref{check1}) and (\ref{check2}) acquire the GSC factor $D_{ph}$ in front of $B_{ph}^{\dagger}(JM)$ on the right-hand sides because, with respect to the correlated ground state $|0\rangle$, relation (\ref{<BB>}) is equivalent to the approximation
\begin{equation}
[B_{ph}(JM),B_{p'h'}^{\dagger}(J'M')]\simeq\delta_{JJ'}\delta_{MM'}\delta_{j_pj_p'}\delta_{j_hj_h'}D_{ph}~,
\label{BB1}
\end{equation}
rather than the approximation (\ref{QBA1}), which is valid only for the HF ground state $|HF\rangle$.

The central result of the phRRPA method is the equation for 
GSC factor $D_{ph}$~\cite{CaDaSa}, which is found by combining Eqs. (\ref{GSCDph}) and (\ref{Rfpfh}) as 
\begin{equation}
D_{ph} \equiv f_h - f_p =1 - \sum_{Ji}(2J+1)\bigg[\frac{1}{2j_p+1}\sum_{h'}D_{ph'}({\cal Y}_{ph'}^{Ji})^2+\frac{1}{2j_h+1}\sum_{p'}D_{p'h}({\cal Y}_{p'h}^{Ji})^2\bigg]~.
\label{Dph}
\end{equation}
The QBA is justified only when the sum on the right-hand side of Eq. (\ref{Dph}) becomes negligible compared to 1, so that $D_{ph}\simeq$ 1. 

The phRRPA equation is obtained in the standard way~\cite{Ring} by using the renormalized phonon operator (\ref{RQ}) and Eq. (\ref{<BB>}) and a model Hamiltonian consisting of a mean field of single-particles with energies $\epsilon_k$ on the spherical orbitals $|j_k,m_k\rangle$ ($k = p, h$) and two-body residual interactions $\langle kk'|V_{res}| ll'\rangle$. The matrix form of the phRRPA equation is given as~\cite{CaDaSa}
\begin{equation}
\left( \begin{array}{cc}
A & B  \\
-B & -A \end{array} \right)\left( \begin{array}{c}
{\cal X}^{Ji} \\
{\cal Y}^{Ji}\end{array} \right)
={E}_{Ji}\left( \begin{array}{c}
{\cal X}^{Ji} \\
{\cal Y}^{Ji}\end{array} \right)~.
\label{RRPA}
\end{equation}
The matrices $A$ and $B$ have the form
\begin{equation}
A_{ph, p'h'} = (\epsilon_p - \epsilon_h)\delta_{pp'}\delta_{hh'} +\sqrt{D_{ph}D_{p'h'}}\langle ph'|V_{res}|hp'\rangle~,\hspace{5mm}
B_{ph, p'h'} = \sqrt{D_{ph}D_{p'h'}}\langle pp'|V_{res}|hh'\rangle~,
\label{AB}
\end{equation}
where the residual interactions are renormalized by the factor $\sqrt{D_{ph}D_{p'h'}}$. Within the QBA, when $D_{ph}=$ 1, one recovers from Eqs. (\ref{RRPA}) and (\ref{AB}) the conventional RPA equation, and from Eqs. (\ref{Rfpfh}) the expressions for occupation numbers $f_k$ (\ref{fpfh}) within RPA.  

The phRRPA equation (\ref{RRPA}) is solved self-consistently with Eq. (\ref{Dph}) and the normalization condition (\ref{norm}). First the RPA matrices ($D_{ph}=$ 1) are diagonalized to determine the eigenvalues ${E}_{Ji}(1)$ (phonon energies), and amplitudes ${\cal X}^{Ji}_{ph}(1)$ and ${\cal Y}^{Ji}_{ph}(1)$. The latter are then used to calculate $D_{ph}(1)$ following Eq. (\ref{Dph}). In the second step, the residual interaction, which is renormalized by $\sqrt{D_{ph}(1)D_{p'h'}(1)}$ are used in diagonalizing Eq. (\ref{RRPA}) (with $\sqrt{D_{ph}(1)D_{p'h'}(1)}$ in the submatrices A and B) to obtain  ${E}_{Ji}(2)$, ${\cal X}^{Ji}_{ph}(2)$, ${\cal Y}^{Ji}_{ph}(2)$. The process is repeated until the convergency is reached, that is 
$|{E}_{Ji}(n) - {E}_{Ji}(n-1)| < 10^{-6}$ MeV. 
\subsection{Transition probability and energy-weighted sum rules}
The nuclear vibrational excitations are generated by the electromagnetic field with one-body excitation operators $\hat{F}_{JM}$ and reduced matrix elements $\langle p||\hat{F}_J||h\rangle$, whose details are given in Refs. \cite{Colo, Bohr}. The reduced transition probabilities $B(EJ, 0\rightarrow \nu)$ between the ground state $|0\rangle$ and one-phonon state $|\nu\rangle\equiv|JMi\rangle$  is calculated within the RRPA as
\begin{equation}
B(EJ, 0\rightarrow \nu) = | \langle\nu |\hat{F}_{J}|0\rangle|^2=\bigg|\sum_{ph}\sqrt{D_{ph}}({\cal X}_{ph}^{Ji} + {\cal Y}_{ph}^{Ji})\langle p ||\hat{F}_J||h\rangle\bigg|^2~.
\label{BE}
\end{equation}
The factor $\sqrt{D_{ph}}$ in Eq. (\ref{BE}) comes from the matrix elements of the transition densities $\rho^{(1)\nu}_{ph}$ and $\rho^{(1)\nu}_{hp}$~\cite{Ring}  within the RRPA 
\[
\rho^{(1)Ji}_{ph} = \langle 0|B_{ph}(JM)|JMi\rangle=\langle 0|B_{ph}(JM){\cal Q}_{JMi}^{\dagger}|0\rangle = 
\sqrt{D_{ph}}{\cal X}_{ph}^{Ji}~,
\]
\begin{equation}
\rho^{(1)Ji}_{hp} = \langle 0|B_{ph}^{\dagger}(JM)|JMi\rangle=\langle 0|B_{ph}^{\dagger}(JM){\cal Q}_{JMi}^{\dagger}|0\rangle = 
\sqrt{D_{ph}}{\cal Y}_{ph}^{Ji}~,
\end{equation}
following the inverse transformation (\ref{RB}) and orthonormal condition (\ref{ortho}), which hold for ${\cal Q}^{\dagger}_{JMi}$ and ${\cal Q}_{JMi}$ with respect to the correlated ground state $|0\rangle$. 

In practice, the distribution of $B(EJ, E_{Ji})$ values of $B(EJ, 0\rightarrow \nu)$ (\ref{BE}) at discrete phonon energies $E_{Ji}$ is also presented as a continuous strength function $S_{J}(E)$ by using the $\delta$-function representation $\delta(x) = \varepsilon/[\pi(x^2+\varepsilon^2)]$, that is
\begin{equation}
S_{J}(E) = \frac{\varepsilon}{\pi}\sum_i\frac{B(EJ, E_{Ji})}{(E-E_{Ji})^2+\varepsilon^2}~.
\label{strength}
\end{equation}
This kind of presentation not only improves the visual comparison of results obtained within the RPA and RRPA, but also mimics the so-called escape width $\Gamma^{\uparrow}\equiv2\varepsilon$, which arises from the direct decay to the continuum, consisting of a free nucleon and a hole state. This escape width $\Gamma^{\uparrow}$ is around few hundred keV, compared to the total GDR width of around 4 - 5 MeV in medium and heavy nuclei.

The energy-weighted sum of rule (EWSR) of the strength distribution generated by the operator $\hat{F}_J$ is defined in terms of its the first moment as 
\begin{equation}
m_1 = \int_0^{E_{max}}ES_J(E)dE = \sum_{i}E_{Ji}B(EJ, E_{Ji})~,
\label{m1}
\end{equation}
This first moment is equal to the half of the expectation value of the double commutator $\langle 0|[\hat{F}_J,[H,\hat{F}_J]]|0\rangle$ if $H$ is the exact two-body Hamiltonian and $|0\rangle$ is the exact ground state~\cite{Ring}. Therefore the EWSR $\langle 0|[\hat{F}_J,[H,\hat{F}_J]]|0\rangle/2$ is also referred to as the double commutator sum rule, which is employed to test the validity of any approach to see if the model-predicted first moment at the left-hand side of Eq. (\ref{m1}) fulfills the EWSR.

For the dipole excitations the IS and IV EWSR, $m_1^{IS}$ and $m_1^{IV}$, are
\begin{equation}
m_1^{IS} = \frac{\hbar^2}{2m^{*}}\frac{A}{4\pi}(33\langle r^4\rangle - 25\langle r^2\rangle^2) ~,\hspace{5mm} m_1^{IV} = (1+\kappa)TRK~,
\label{EWSR}
\end{equation}
respectively, where 
\begin{equation}
TRK = \frac{9}{4\pi}\frac{\hbar^2}{2m^{*}} \frac{NZ}{A}~
\label{TRK}
\end{equation}
is the model independent Thomas-Reiche-Kuhn (TRK) sum rule, and $\kappa$ is the enhancement factor owing to the velocity dependence and exchange mixtures (Wigner force) in the residual interaction. This enhancement factor $\kappa$ is calculated according to Eq. (35) of Ref. \cite{Colo}. In Eqs. (\ref{EWSR}) and (\ref{TRK}) the corrected mass $m^{*}\equiv m A/(A-1)$ is used instead of the bare nucleon mass $m$ in all Skyrme-HF calculations to take care of the center-of-mass effect on the total energy. 
\section{The RRPA including the effect of particle-particle and hole-hole correlations}
\label{fRRPA}
As has  been mentioned in the Introduction, all extensions of the RPA in the spirit of phRRPA so far violate the sum rules. The reason comes from the presence of the GSC factors $D_{ph}$ in the transition strengths \eqref{BE}, which are reduced because $D_{ph}<$ 1. This major drawback of the phRRPA was noticed for the first time in Ref.  \cite{Hirsch}, where it has been pointed out that the renormalized quasiparticle RPA (RQRPA) violates the Ikeda sum rule for the charge exchange excitations such as the Gamow-Teller resonance or for the Fermi transitions between the ground state of the initial even-even nuclei and the ground as well as excited states of the odd-odd nuclei. For the dipole excitations, the GSC factor $D_{ph}$ causes the decrease of the EWSRs. To resolve this problem, one needs to take into account the contribution owing to the expectation value of the commutation relation 
\begin{equation}
\langle{0}|[B_{ss'}, B^{\dagger}_{s'_1s_1}]|{0}\rangle\simeq \delta_{ss_1}\delta_{s's'_1}D_{ss'} ~,
\label{Dss}
\end{equation}
which is neglected within the phRRPA. In Eq. \eqref{Dss}, $D_{ss'} = f_{s'} - f_{s}$ ($ss' = pp'$ or $hh'$) are the correlation factors coming from the pp and hh channels. As the result, the total transition probabilities $B(EJ)$ are the sum of the those associated with the ph, pp, and hh excitations. These transition probabilities are denoted as $B^{ph}(EJ)$, $B^{pp'}(EJ)$, and $B^{hh'}(EJ)$, respectively.
\begin{equation}
B(EJ, 0\rightarrow \nu) = B^{ph}(EJ, 0\rightarrow \nu) + B^{pp'}(EJ, 0\rightarrow \nu) + B^{hh'}(EJ, 0\rightarrow \nu) ~, \label{BEJ}
\end{equation}
where
\begin{equation}
B^{ph}(EJ, 0\rightarrow \nu) =\bigg|\sum_{ph}\sqrt{D_{ph}}({\cal X}_{ph}^{Ji} + {\cal Y}_{ph}^{Ji})\langle p ||\hat{F}_J||h\rangle\bigg|^2 ~,
\label{BEJph}
\end{equation}
\begin{equation}
B^{pp'}(EJ, 0\rightarrow \nu) =\bigg|\sum_{pp'}\sqrt{D_{pp'}}({\cal X}_{pp'}^{Ji} + {\cal Y}_{pp'}^{Ji})\langle p ||\hat{F}_J||p'\rangle\bigg|^2 ~,
\label{BEJpp}
\end{equation}
\begin{equation}
B^{hh'}(EJ, 0\rightarrow \nu) =\bigg|\sum_{hh'}\sqrt{D_{hh'}}({\cal X}_{hh'}^{Ji} + {\cal Y}_{hh'}^{Ji})\langle h ||\hat{F}_J||h'\rangle\bigg|^2  ~,
\label{BEJhh}
\end{equation}
with ${\cal X}_{ss'}^{Ji}$ and ${\cal Y}_{ss'}^{Ji}$ being the vibration amplitudes that correspond to the pp and hh excitations, whereas the GSC factors  $D_{pp'}$ and $D_{hh'}$ are expressed explicitly in terms of the GSC $D_{ph}$ and the amplitudes ${\cal Y}_{ph}^{Ji}$ by using Eqs. (\ref{Rfpfh}) and (\ref{Dss})~\cite{ERPA} as
\begin{equation}
D_{pp'} \equiv f_{p'} - f_p = \sum_{Ji}(2J+1)\sum_{h}\bigg[\frac{1}{2j_{p'}+1}D_{p'h}({\cal Y}_{p'h}^{Ji})^2-\frac{1}{2j_p+1}D_{ph}({\cal Y}_{ph}^{Ji})^2\bigg]~,
\label{Dpp}
\end{equation}
\begin{equation}
D_{hh'} \equiv f_{h'} - f_h = \sum_{Ji}(2J+1)\sum_{p}\bigg[\frac{1}{2j_{h'}+1}D_{ph'}({\cal Y}_{ph'}^{Ji})^2-\frac{1}{2j_h+1}D_{ph}({\cal Y}_{ph}^{Ji})^2\bigg]~.
\label{Dhh}
\end{equation}

In principle, the phonon amplitudes ${\cal X}_{ss'}^{Ji}$ and ${\cal Y}_{ss'}^{Ji}$ should be obtained by diagonalizing the full RRPA matrix, including all the ph, pp, and hh configurations, similar that obtained in Ref. \cite{Sommer} for the FT-RPA. 
However, doing so will significantly enlarge the size of the RRPA matrix, leading to a sharp increase of computational time. This project is now underway. In the present study, for the feasibility in practical calculations as a preparatory step to test the effect of GSC in the restoration of the GDR sum rule, we employ the extended RRPA (ERRPA), proposed in Ref.~\cite{ERPA}, which considers a model Hamiltonian with a separable residual interaction in the form of Eq. (7) in Ref.~\cite{ERPA}. The full RRPA matrix equations with the separable interaction can be easily transformed to obtain the phonon amplitudes in an explicit form, including the ${\cal X}^{Ji}_{ss'}$ and ${\cal Y}^{Ji}_{ss'}$ of the pp and hh excitations (Eqs. (18) -- (20) in Ref. \cite{ERPA}). All the eigenvalues of the full RRPA equations are obtained by solving the secular equation (21) therein, instead of diagonalizing the full RRPA matrix. 

As the formalism proposed on Ref. \cite{ERPA} is derived only for separable interactions, to use it in our study, we adopt the approximate factorization of the interaction~\cite{Ring}
\begin{equation}
\langle  \lefteqn{\overbrace{\phantom{kl|V_{res}|l}}^J}k\underbrace{l|V_{res}|lk}_J \rangle ~\simeq\lambda^{(J)}f_{kl}^2~,\hspace{5mm} (kl = ph, pp', hh')~,
\label{separable}
\end{equation}
with $\lambda^{(J)}=$1. Because the contribution of $pp$ and $hh$ excitations is expected to be small, we treat it in a perturbative way. To this end, instead of solving Eq. (21) of Ref. \cite{ERPA} to obtain all the eigenvalues $E_{Ji}^{ph}$, $E_{Ji}^{pp'}$, and $E_{Ji}^{hh'}$ as well as the amplitudes ${\cal X}_{ph}^{Ji}$, ${\cal Y}_{ph}^{Ji}$, ${\cal X}_{ss'}^{Ji}$ and ${\cal Y}_{ss'}^{Ji}$, we approximate the energies $E^{Ji}_{ph}$ and the amplitudes ${\cal X}_{ph}^{Ji}$ and ${\cal Y}_{ph}^{Ji}$ with the corresponding values obtained within the phRRPA described in Sec. \ref{secRRPA}. Regarding the energies of the new phonon states, which appears between the poles $\epsilon_s - \epsilon_{s'}$ of pp and hh excitations, their energies $E_{Ji}^{ss'}$ are considered to be close to the corresponding poles $\epsilon_s - \epsilon_{s'}$, namely 
\begin{equation}
E_{Ji}^{ss'} = \epsilon_s - \epsilon_{s'} - \delta E~.
\label{dE}
\end{equation}
The energy shift $\delta E$ is adjusted to restore the EWSR, which is violated within the phRRPA. In this way, although $\delta E$ plays the role of a parameter of the model, it has a physical justification as the difference between the new RRPA solutions due to the pp (hh) configurations and the corresponding pp (hh) poles. The amplitudes ${\cal X}_{ss'}^{Ji}$ and ${\cal Y}_{ss'}^{Ji}$ are calculated by using the expression obtained with the separable interaction given by Eqs. (19) and (20) of Ref. \cite{ERPA} with $k=1$, whereas $f_{ph}$ and $f_{ss'}$ are approximated with $\sqrt{|\langle|ph|V_{res}|hp\rangle|}$ and $\sqrt{|\langle ss'|V_{res}|s's\rangle|}$, respectively, according to Eq. (\ref{separable}). This version of phRRPA, which takes into account the contribution of pp and hh excitations, is referred to as the RRPA hereafter. 

The advantage of this approximation is its simplicity, based on the solutions of the phRRPA, avoiding the diagonalization of a large-size matrix.
Its shortcoming is the lost of self-consistency, which may not be serious so long as the GSC is not large to justify the validity of the perturbative approximation employed here.  
 
\section{Analysis of the numerical results}
\label{results}
The computer code for the self-consistent HF-RPA with Skyrme-type interactions, developed by Col\`{o} and collaborators~\cite{Colo}, is modified to include the effect of GSC beyond RPA. This code is restricted to the calculations of natural-parity states  having $\pi = (-)^{J}$ in spherical nuclei with filled subshells, that is without partial occupancies. The single-particle energy spectra for neutrons and protons are discretized and superfluid pairing is not included. Regarding the effective nucleon-nucleon interaction, the density-dependent Skyrme interactions are employed in the calculations. The code first carries out the HF calculations for a specified nucleus using a given  Skyrme-type interaction. The HF single particle energies, wave functions and densities, which are obtained in a radial mesh extending to $R$ (fm)  and cut-off energy  $E_c$ of unoccupied states, are then used in solving the RPA equation. As for the phRRPA and RRPA equations, they are solved by iteration as has been specified respectively in Secs. \ref{secRRPA} and \ref{fRRPA} above. For the details of the HF-RPA code see Ref. \cite{Colo}.

In the present paper we restrict ourselves to studying the IS and IV dipole states in $^{48, 52, 58}$Ca and $^{90, 96, 110}$Zr isotopes. For the Skyrme interactions, we employ the SLy5 force, whose parameters have been specified and used in the HF-RPA code in a spherical box with radius $R=$ 15 fm and the cut-off energy $E_c=$ 60 MeV of unoccupied states. This value has been chosen because, for the IV E1 excitations in the nuclei under consideration, the energy of the lowest RPA state decreases very slowly with increasing $E_c$ higher than 60 MeV. A value $\varepsilon=$ 0.4 MeV for the smoothing parameter is used in calculating the strength function (\ref{strength}).
    \begin{figure}
       \includegraphics[height=17cm]{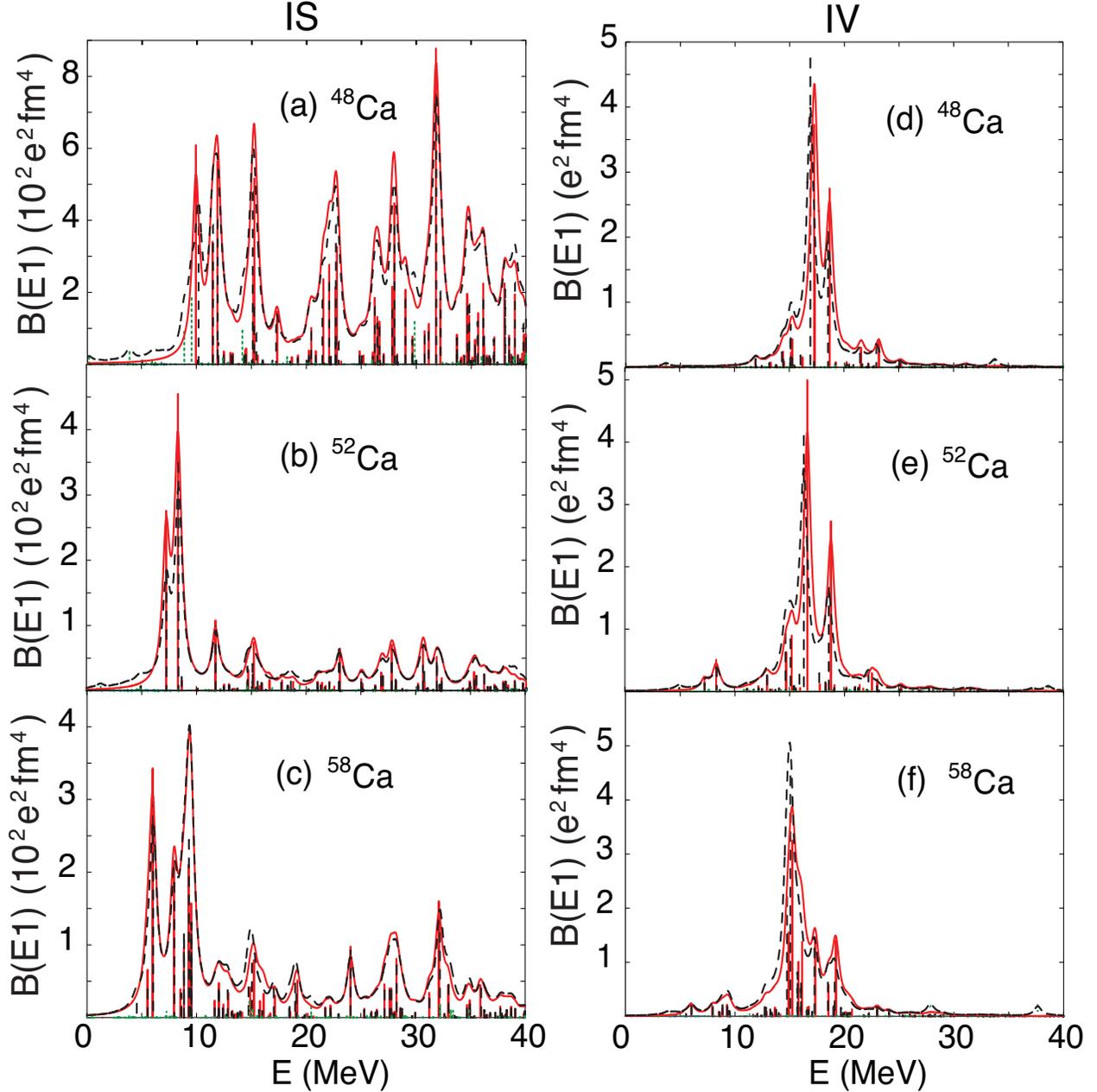}
       \caption{(Color online) Distributions of the IS (left panels) and IV (right panels) reduced transition probabilities $B(E1, 0 \rightarrow 1^-)$ and the corresponding strength functions $S(E)$ for $^{48}$Ca [(a) and (d)], $^{52}$Ca [(b) and (e)], and $^{58}$Ca [(c) and (f)] obtained within the RPA, phRRPA, and RRPA. The solid and dashed vertical bars denote the $B(E1)$ values obtained within the RPA and phRRPA, respectively. The dotted vertical bars stand for the $B^{pp'}(E1)$ calculated based on Eq. \eqref{BEJpp}. The continuous solid and dashed lines depict the total strength functions $S(E)$ obtained within the RPA and RRPA [Eq. \eqref{BEJ}], respectively. The units on the $y$-axis stand for the $B(E1, 0 \rightarrow 1^-)$ values, whereas the units of the strength function $S(E)$ are equal to those of $B(E1, 0 \rightarrow 1^-)$ divided by MeV.
        \label{BE1Ca}}
    \end{figure}
    \begin{figure}
       \includegraphics[width=18cm]{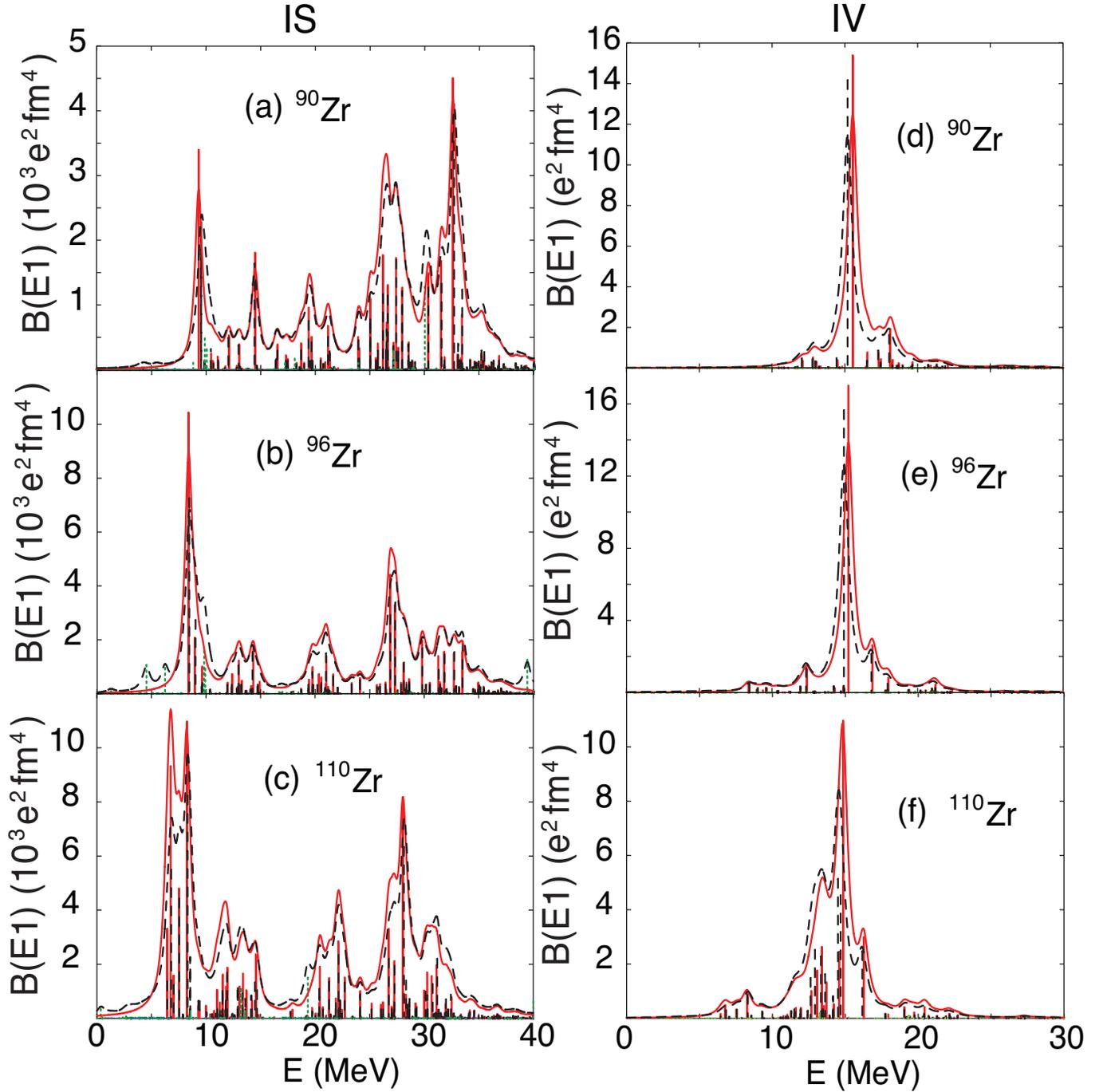}
       \caption{(Color online) The same as in Fig. \ref{BE1Ca} but for $^{90, 96, 110}$Zr isotopes. 
        \label{BE1Zr}}
    \end{figure}
\begin{table}
\caption{The fulfillment of the EWSR for the IS and IV within the RPA, phRRPA, and RRPA for the nuclei under consideration in the paper. The results are obtained by using the SLy5 interaction with the summation taken from 0 to $E_{max}=$ 60 MeV.}
\begin{tabular}{|c|c|c|c|c|c|c|c|c|c|c|c|c|}
\hline\hline
&\multicolumn{4}{|c|}{IS} & \multicolumn{4}{c|}{IV} \\
\hline
	&$m_1$&${m_1^{RPA}}/{m_1}$&${m_1^{RRPA}}/{m_1}$&${m_1^{phRRPA}}/{m_1}$&$TRK$&${m_1^{RPA}}/{TRK}$&${m_1^{RRPA}}/{TRK}$&${m_1^{phRRPA}}/{TRK}$\\ &$\times 10^5$&$(\%)$&$(\%)$ & $(\%)$&&&&\\
\hline
    $^{48}$Ca&2.52& 99.80 & 100.08 & 91.16 & 196.6 & 1.17 & 1.09 & 1.02 \\
    $^{52}$Ca&3.47& 99.80 & 100.40 & 86.98 & 179.3 & 1.17 & 1.07 & 0.97 \\
   $^{58}$Ca&4.57& 99.85 & 99.44 & 88.79 & 191.2 & 1.17 & 1.12 & 0.97 \\
   $^{90}$Zr&9.05& 100.03 & 100.98 & 92.00 & 326.3 & 1.18 & 1.04 & 1.03 \\
   $^{96}$Zr&11.2& 100.04 & 100.68 & 90.82 & 342.9 & 1.18 & 1.11 & 1.04 \\
  $^{110}$Zr&16.4& 99.95 & 100.95 & 88.89 & 374.6 & 1.18 & 1.05 & 1.01 \\
    \hline\hline
\end{tabular}
\label{table1}
\end{table}
\begin{table}
\caption{Centroid energies calculated from the IS and IV $B(E1, 0 \rightarrow 1^-)$ strength distributions obtained within the RPA, phRRPA, and RRPA for the nuclei under consideration in the paper. The last column presents the results calculated from the empirical fitting of the experimental centroid energy, namely $E_m=31.2A^{-1/3}+20.6A^{-1/6}$ \cite{Em}.}
\begin{tabular}{|c|c|c|c|c|c|c|c|c|c|c|c|}
\hline\hline
&\multicolumn{3}{|c|}{IS} & \multicolumn{4}{c|}{IV} \\
\hline
	&RPA&phRRPA&RRPA&RPA&phRRPA&RRPA&$E_m$\\
\hline
    $^{48}$Ca&26.60 & 26.87 & 29.74 & 18.22 & 17.67 & 18.24 &19.39\\
    $^{52}$Ca&18.95 & 20.10 & 23.20 & 17.17 & 16.67 & 18.40 & 19.02\\
    $^{58}$Ca&17.56 & 18.39 & 20.59 & 16.18 & 15.70 & 18.02 & 18.53\\
    $^{90}$Zr&25.40 & 26.08 & 28.63 & 16.28 & 15.83 & 15.99 & 17.16\\
    $^{96}$Zr&21.49 & 22.31 & 24.73 & 15.54 & 15.18 & 16.24 & 16.44\\
   $^{110}$Zr&18.07 & 19.08 & 21.67 & 14.40 & 14.04 & 14.60 & 15.92\\
    \hline\hline
\end{tabular}
\label{table2}
\end{table}
\begin{table}
\caption{Ratio $r = S_{PDR}/S_{GDR}(\%)$ obtained within the RPA, phRRPA, and RRPA for calcium and zirconium isotopes.}
\begin{tabular}{|c|c|c|c|}
\hline\hline
&~RPA~&~phRRPA~&~RRPA~ \\
\hline
	$^{48}$Ca&0.02&0.02&0.33 \\
	$^{52}$Ca&2.82&2.67&3.24 \\
	$^{58}$Ca&3.99&4.20&4.20 \\
	$^{90}$Zr&0.01&0.01&0.06 \\
	$^{96}$Zr&2.37&2.44&3.02 \\
	$^{110}$Zr&4.06&4.13&4.24 \\
\hline\hline
\end{tabular}
\label{table3}
\end{table}

Shown in Figs. \ref{BE1Ca} and \ref{BE1Zr} are the IS and IV dipole distributions in calcium and zirconium isotopes obtained within the RPA, phRRPA, and RRPA. The values of the energy shift $\delta E$ in Eq. (\ref{dE}) are chosen equal to 1 MeV for calcium isotopes, 1.5 MeV for both $^{90}$Zr and $^{96}$Zr, and 1.4 MeV for $^{110}$Zr. It is seen from these figures that, as compared to the RPA results, the overall IS strengths obtained within the phRRPA are always smaller, whereas the phRRPA IV strength distribution is  slightly shifted to the lower excitation energy with an increase in the total strength at the low-energy side of the GDR main peak and a depletion strength at its high-energy side. As the result of these shifts in the distributions of the IS and IV strengths, the EWSRs calculated within the phRRPA are significantly reduced,  as shown in the 5$^{th}$ and 9$^{th}$ columns of Table \ref{table1}. For the IS excitations, the largest value of $m_1^{phRRPA}/m_1$ amounts to only 92.0$\%$ for $^{90}$Zr, whereas the smallest value is 86.98$\%$ for $^{52}$Ca. For the IV excitations, the value of $m_1^{phRRPA}/TRK$ ranges from 0.97 to 1.04. 

The first excited IS state, which is the spurious mode caused by the center-of-mass motion, is shifted up within the phRRPA. However, in all the cases under consideration, this increase in the energy of the spurious mode is not large and the associated $B(E1)$ values are negligible. Similar to the RPA case, the phRRPA spurious mode is still well separated from all physical states located at higher energies, and therefore does not affect these states. Within the RRPA, there are several IS and IV transition strengths caused by the pp excitations (green dotted vertical bars), but only few enhanced strengths are seen in the IS distributions, especially in $^{48}$Ca and $^{90,96}$Zr, whereas those seen in the IV distribution are quite small. The transitions associated with hh excitations are found to be negligible in the present calculations and, therefore, not shown in the figures. The largest IV transition strengths associated with the pp excitations are seen in the low energy (PDR) region, which slightly redistributes  the PDR strengths, especially for calcium isotopes. Adding the transition strengths associated with the pp excitations, the EWSRs obtained within the RRPA are fully recovered for both the IS and IV excitations as seen in the 4$^{th}$ and 8$^{th}$ columns of Table \ref{table1}, respectively. This result is particularly interesting because it reveals the reason why all extensions of RPA so far violate the EWSRs: They have neglected the transitions associated with the pp and hh excitations.

To have a deeper look inside the effects of GSC beyond the RPA, we report in Table \ref{table2} the centroid energies $\bar{E}$, which are defined as $\bar{E}=m_1/m_0$, where $m_1$ is calculated from Eq. \eqref{m1} and $m_0=\sum_i B(EJ, E_{Ji})$, obtained within the RPA, phRRPA, and RRPA for the IS and IV strength distributions. For the IS mode, as compared to the RPA, $\bar{E}$ obtained within the phRRPA increases slightly, whereas a significant increase in $\bar{E}$ obtained within the RRPA is observed, implying the significant contribution of the pp and hh strengths. For the IV mode, $\bar{E}$ obtained within the phRRPA is always lower than that obtained within the RPA, whereas the RRPA centroid energy increases to be closer to the data obtained from the empirical fitting of the experimental centroid energy \cite{Em} as compared to the RPA one, except for $^{90}$Zr nucleus. Once again, this increase in the centroid energy obtained within the RRPA for the IV mode shows the important contribution of the pp and hh transitions.

To see the contribution of the PDR to the total $B(E1)$ strength, we present in Table \ref{table3} the ratio $r = S_{PDR}/S_{GDR} (\%)$ of the energy weighted sum of strength (EWSS) of the PDR to that of the GDR, where $S = \sum_\nu{E_\nu B(EJ, E_\nu)}$. The summation is taken within 0$\leq E_\nu\leq$ 10 MeV for the PDR and  0$\leq E_\nu\leq$ 60 MeV for the GDR. The values of $r$ increase from almost 0$\%$ in the stable nuclei, such as $^{48}$Ca and $^{90}$Zr, to about 4$\%$ in very neutron-rich nuclei, such as $^{58}$Ca and $^{110}$Zr. As compared to the RPA, the phRRPA increases the ratio $r$ in most nuclei under consideration, except in $^{52}$Ca, whereas this ratio is significantly enhanced within the RRPA due to the contributions of the pp transition strengths in the PDR region.
\section{Conclusions}
The present work studies the effects of GSC beyond RPA on IS and IV strength distributions in calcium and zirconium isotopes. The self-consistent RPA code using  Skyrme interactions~\cite{Colo} has been modified and applied in the numerical calculations to include the effects of GSC beyond RPA within the phRRPA~\cite{CaDaSa}. To restore the EWSR, which is violated within the phRRPA, we propose a simple RRPA calculations taking into account, in addition to the ph excitations, the contribution of the pp and hh excitations in a perturbative way, which increases the total energy-weighted sum of strengths, including those from the ph ones.

The analysis of numerical calculations obtained by using the SLy5 interactions allows us to draw the following conclusions.

1 - Although GSC beyond RPA shift up the energy of the spurious mode obtained within the phRRPA and increases the corresponding B(E1) value for the IS spurious transition, this increase however is not large and the phRRPA spurious mode is still well separated from all physical states.

2 - GSC beyond RPA reduces the transition strengths associated with the IS mode, whereas it slightly increases the total strength on the low-energy region (the PDR region) and decreases the strength on the other side (the GDR region) leading to a significant decrease of the EWSRs for both the IS and IV modes obtained within the phRRPA. This violation of the sum rule is then fully recovered by taking into account the contribution of pp and hh excitations within the RRPA. This result reveals the reason why all the RPA extensions that do not take into account the pp and hh excitations violate the EWSRs.

3 - The ratio of the PDR EWSS to the GDR one, which is almost zero in stable nuclei, increases with the neutron number. As compared to the RPA case, this ratio is in general significantly larger within the RRPA.

In the present RRPA calculations, for feasibility and simplicity of numerical calculations, the contribution of the pp and hh excitations is considered approximately in an perturbative way. To perform the complete RRPA calculations, we need to employ the method proposed in Ref. \cite{Sommer} or \cite{ERPA} in a fully self-consistent way, as has been mentioned previously. Moreover, the exact superfluid pairing, which has been shown to have a non-negligible effect on the PDR~\cite{pairing}, is neglected in the present study. The extension of the HF mean-field approach incorporated with the RRPA into the renormalized quasiparticle RPA, taking into account the pp and hh excitations together with exact pairing, remains one of the goals in our future study. 

\acknowledgments
The numerical calculations were carried out using the Supermicro Intel Xeon E5-2670 Server of the Institute de Physique Nucl\'{e}aire (IPN), Orsay (France) and the HOKUSAI-GreatWave supercomputer system at RIKEN.

The authors are grateful to G. Col\`{o} and L. Cao for the fruitful discussions related to the use of the computer code for self-consistent RPA with Skyrme-type interactions, and P. Quentin for reading the manuscript and useful comments.

N.Q.H. acknowledges the support by the RIKEN Interdisciplinary Theoretical Science (iTHES) Research Group during his visit in RIKEN, where the initial part of this work was carried out. N.Q.H. is thankful to the support by the National Foundation for Science and Technology Development (NAFOSTED) of Vietnam through Grant No. 103.04-2013.08. T.V.N.H acknowledges the support by the U.S. NSF Grant No. 1415656 and U.S. DOE Grant No. DE-FG02-08ER41533.


\begin{thebibliography}{99}
\bibitem{Ring}P. Ring and P. Schuck, {\it The Nuclear Many-Problem} (Springer, 2004).
\bibitem{Hara}K. Hara, Prog. Theor. Phys. {\bf 32},  88 (1964); K. Ikeda, T. Udagawa, and H. Yamaura, Prof. Theor. Phys. {\bf 33}, 22 (1965).
\bibitem{CaDaSa}F. Catara, N. Dinh Dang, and M. Sambataro, Nucl. Phys. A {\bf 579}, 1 (1994).
\bibitem{Dukelsky}J. Dukelsky and P. Schuck, Phys. Lett. B {\bf 464}, 164 (1999).
\bibitem{Rowe}D.J. Rowe, Phys. Rev. {\bf 175}, 1283 (1968).
\bibitem{Schuck}P. Schuck and S. Ethofer, Nucl. Phys. A {\bf 212}, 269 (1973).
\bibitem{Lenske}H. Lens and J. Wambach, Phys. Lett. B {\bf 249}, 377 (1990).
\bibitem{Catara}F. Catara, G. Piccitto, M. Sambataro, and N. Van Giai, Phys. Rev. B {\bf 54}, 17536 (1996).
\bibitem{Liu}K.F. Liu, Phys. Lett. B{\bf 60}, 9 (1975).
\bibitem{Hirsch}J.G. Hirsch, P.O. Hess, and O. Civitarese, Phys. Rev. C {\bf 54}, 1976 (1996).
\bibitem{Ignatyuk}A.V. Igantyuk, Sov. J. Nucl. Phys. {\bf 21}, 10 (1975).
\bibitem{Sommer} H. M. Sommermann, Ann. Phys. {\bf 151}, 163 (1983).
\bibitem{Vautherin}D. Vautherin and N. Vinh Mau, Nucl. Phys. A {\bf 422}, 140 (1984). 
\bibitem{DangJP}N. Dinh Dang, J. Phys. G {\bf 11}, L125 (1985).
\bibitem{Lacroix}D. Lacroix, Ph. Chomaz, and S. Ayik, Phys. Rev. C {\bf 58}, 2154 (1998).
\bibitem{PDRexp} A. Liestenscheneider {\it et al.}, Phys. Rev. Lett. {\bf 86}, 5442 (2001); E. Tryggestad
{\it et al.}, Phys. Lett. B {\bf 541}, 52 (2002); T. Aumann {\it et al.}, Eur. Phys. J. A {\bf 26}, 441 (2005); O. Wieland and A. Bracco, Prog. Part. Nucl. Phys. {\bf 66}, 374 (2011). 
\bibitem{ReRPA} D. Vretenar, N. Paar, P. Ring, and G.A. Lalazissis, Nucl. Phys. A {\bf 692}, 496 (2001); N. Paar,
T. Niksic, D. Vretenar, and P. Ring, Phys. Lett. B {\bf 606}, 288 (2005).
\bibitem{Tarpanov} D. Tarpanov, C. Stoyanov, N. Van Giai, and V. V. Voronov, Phys. Atom. Nucl. {\bf 70}, 1402 (2007).
\bibitem{Sarchi}D. Sarchi, P.F. Bortignon, and G. Col\`{o}, Phys. Lett. B {\bf 601}, 27 (2004). 
\bibitem{SRPA} D. Gambacurta, M. Grasso, and F. Catara, Phys. Rev. C {\bf 84}, 034301 (2011). 
\bibitem{ReRPA1} E. Litvinova, P. Ring, D. Vretenar, Phys. Lett. B {\bf 647}, 111 (2007).
\bibitem{QPM1} N.N. Arsenyev, A.P. Severyukhin, V.V. Voronov, and N. Van Giai, Acta Physica Polonica B {\bf 46}, 517 (2015).
\bibitem{QPM2} N. Tsoneva and H. Lenske, Phys. Rev. C {\bf 77}, 024321 (2008).
\bibitem{Gamba}D. Gambacurta and F. Catara, Phys. Rev. B {\bf 77}, 205434 (2008).
\bibitem{Colo}G. Col\`{o}, L. Cao, N. Van Giai, and L. Capelli, Com. Phys. Com. {\bf 184}, 142 (2013).
\bibitem{Hage}M. Hage-Hassan and M. Lambert, Nucl. Phys. A {\bf 188}, 545 (1972).
\bibitem{Ellis}P.J. Ellis, Nucl. Phys. A {\bf 155}, 625 (1970).
\bibitem{Bohr}A. Bohr and B. Mottelson, {\it Nuclear Structure}, Vol. I (Benjamin, 1969).
\bibitem{ERPA} N.D. Dang and A. Arima, Phys. Rev. C {\bf 62}, 024303 (2000).
\bibitem{Em} B.L. Berman and S.C. Fultz, Rev. Mod. Phys. {\bf 47}, 713 (1975).
\bibitem{pairing}N. Dinh Dang and N. Quang Hung, J. Phys. G {\bf 40}, 105103 (2013).

\end{thebibliography}
\end{document}